\documentclass[conference]{IEEEtran}
\IEEEoverridecommandlockouts
\usepackage{cite}
\usepackage{amsmath,amssymb,amsfonts}
\usepackage{algorithmic}
\usepackage{graphicx}
\usepackage{textcomp}
\usepackage{xcolor}
\usepackage[subrefformat=parens]{subcaption}
\usepackage{comment}
\def\BibTeX{{\rm B\kern-.05em{\sc i\kern-.025em b}\kern-.08em
    T\kern-.1667em\lower.7ex\hbox{E}\kern-.125emX}}
\begin{document}

\title{MagGlove: A Haptic Glove with Movable Magnetic Force for Manipulation Learning\\
{\footnotesize \textsuperscript{}}
}

\author{\IEEEauthorblockN{Mikiya Kusunoki}
\IEEEauthorblockA{
\textit{Japan Advanced Institute of}\\
\textit{Science and Technology}\\
Ishikawa, Japan \\
s2210058@jaist.ac.jp}
\and
\IEEEauthorblockN{Shogo Yoshida}
\IEEEauthorblockA{
\textit{Japan Advanced Institute of}\\
\textit{Science and Technology}\\
Ishikawa, Japan \\
s2020047@jaist.ac.jp}
\and
\IEEEauthorblockN{Haoran Xie}
\IEEEauthorblockA{
\textit{Japan Advanced Institute of}\\
\textit{Science and Technology}\\
Ishikawa, Japan \\
xie@jaist.ac.jp}
\and

}

\maketitle

\begin{abstract}
Recently, haptic gloves have been extensively explored for various practical applications, such as manipulation learning.  Previous glove devices have different force-driven systems, such as shape memory alloys, servo motors and pneumatic actuators; however, these proposed devices may have difficulty in fast finger movement, easy reproduction, and safety issues. In this study, we propose MagGlove, a
novel haptic glove with a movable magnet mechanism that has a linear motor, to solve these issues. The proposed MagGlove device is
a compact system on the back of the wearer's hand with high responsiveness, ease of use, and good safety. The proposed device is adaptive with the modification of the magnitude of the current flowing through the coil. Based on our evaluation study,  it is verified that the proposed device can achieve finger motion in the given tasks. Therefore, MagGlove can provide flexible support tailored to the wearers' learning levels in manipulation learning tasks.
\end{abstract}


\section{Introduction}
In recent years, many studies have been conducted to extend human capabilities and senses, known as human augmentation. Wearable devices for physical augmentation have a shape-changing interface for temperature sensing~\cite{xcloth20} and a wearable robot arm as third arm to support the daily activities~\cite{xlimb}. NaviChoker~\cite{navichoker21} uses a pneumatic actuator for the haptic guidance of walking direction. Similar to these human augmentation devices, we propose a sensory augmentation device to guide the wearers' finger movements through a haptic glove. 

\begin{figure}[t]
\begin{center}
\includegraphics[width=0.8\linewidth]{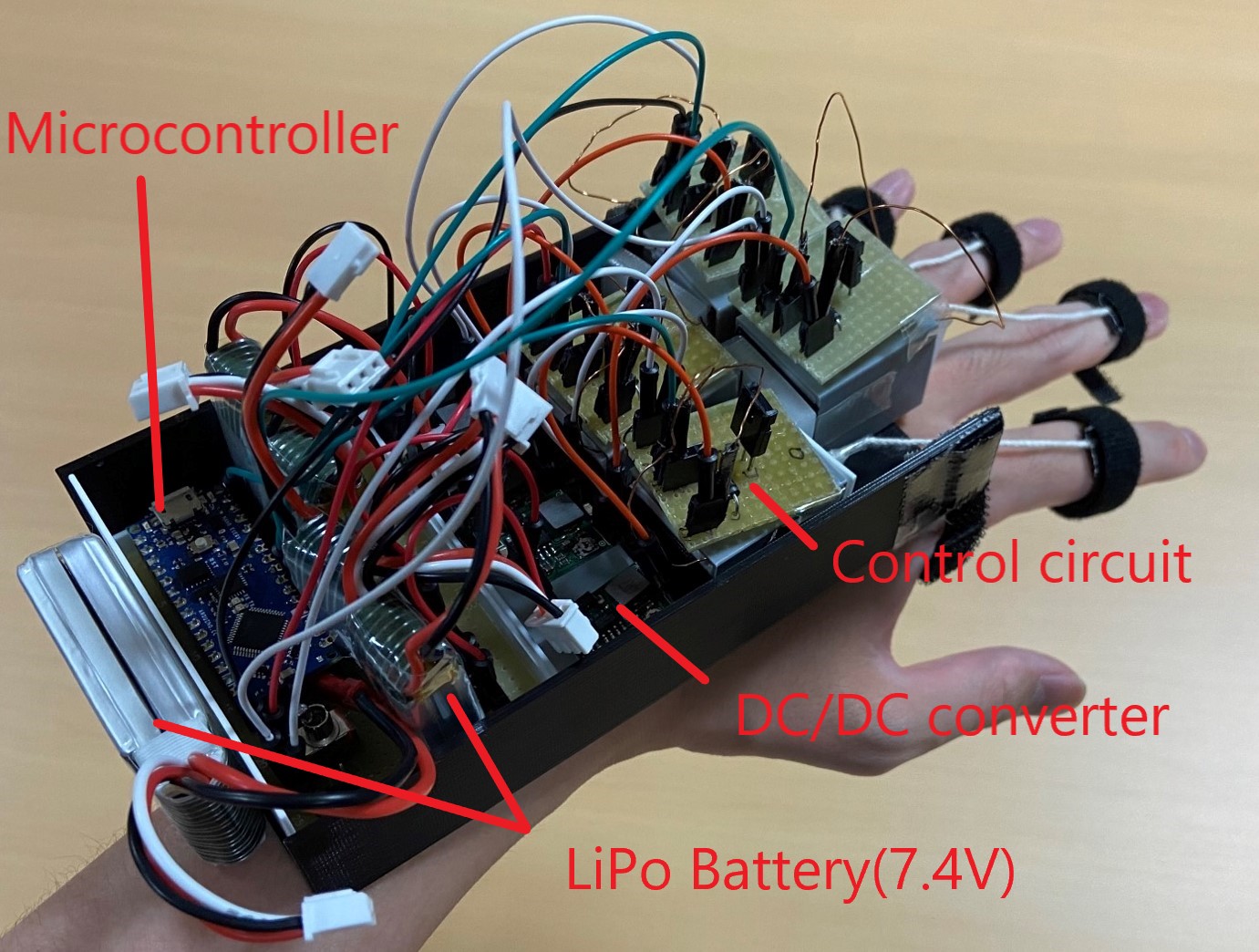}
\label{fig1.1}
\end{center}
\caption{The proposed haptic glove device MagGlove.}
\end{figure}

Robotic gloves utilize various type of actuators. A device that has a shape memory alloy driven~\cite{10.1145/3428361.3428404} may be not applicable to a case where the finger needs to be moved quickly due to its low responsiveness. Devices with a pneumatic actuator~\cite{7139597} may be difficult to use because they require a control valve. The method using servo motors is not safe. Tactiles is a robotic glove that uses electromagnetic actuators~\cite{8797921} to reproduce haptics in virtual reality (VR), but not applicable for manipulation learning. Most existing robotic gloves are designed for rehabilitation purposes for specific users. However, we aim to propose a haptic device only for rehabilitation purposes but also for educational purposes for normal users.

To solve these issues, we propose MagGlove, a haptic glove used to guide the user's finger movements with a wire-driven linear actuator.  Manipulation learning can be conducted using the proposed device, and the users can sense the guiding force and move their fingers. For the proposed MagGlove device, a finger is pulled by the string attached to a magnet actuator. While using this device, the user can grasp the finger to be moved accordingly. In addition, the user can distinguish between finger extension and finger flexion by setting different time intervals for applying force to the finger. We propose a linear actuator with a movable magnetic force, which consists of a coil and a permanent magnet in a simple structure. Due to its independent structure, the wearer can ignore the existence of cables when using this device. The force guidance for the user's fingers with the proposed haptic glove can be used in various applications, such as entertainment and education, and the proposed device is promising in practicing fingerspelling and playing musical instruments.

\section{Proposed Device}

The proposed haptic device MagGlove is shown in Figure 1. The actuator is based on the principle of a linear motor, which uses a wire-driven system for haptic feedback (Figure 2). The coil is made of enameled wire with a diameter of 0.4 mm and a length of 20 m. This enameled wire is wound around a cylinder with a diameter of 15 mm and a height of 32 mm, which is fabricated by a 3D printer (poly-lactic acid resin), as shown in Figure 3(a) and (b). Two neodymium magnets with a diameter of 13 mm, a thickness of 2 mm, and a magnetic flux density of 240 mT are stacked on top of each other and fixed to the top and bottom of a 13 mm diameter and 5 mm high cylinder made by a 3D printer using double-sided tape, as shown in Figure 3(c). The magnets on both ends of the cylinder are fixed so that their poles face each other. The size of the actuator is $25mm \times 25mm \times 32mm$.

\begin{figure}[b]
\begin{center}
\includegraphics[width=0.8\linewidth]{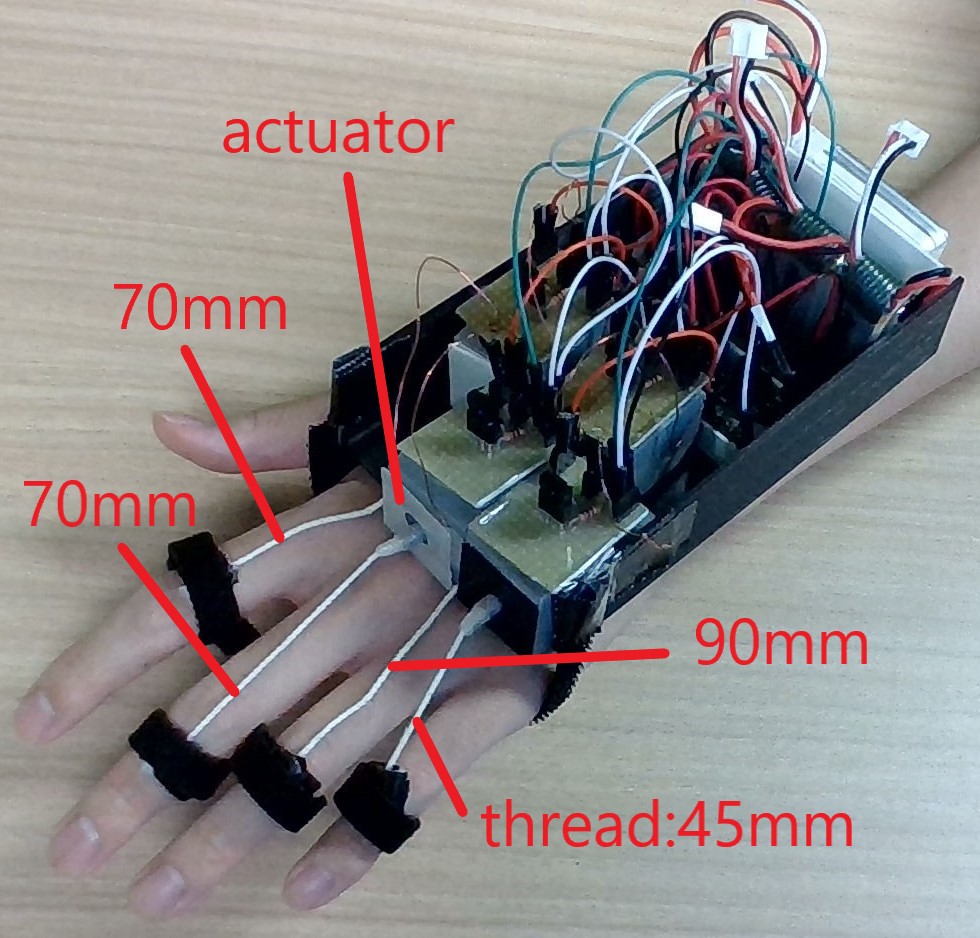}
\caption{The wire-driven actuator system of the proposed device.}
\label{fig1.2}
\end{center}
\end{figure}

\begin{figure}[t]
    \begin{tabular}{cc}
      \begin{minipage}[t]{0.45\hsize}
        \centering
        \includegraphics[height=30mm,width=40mm]{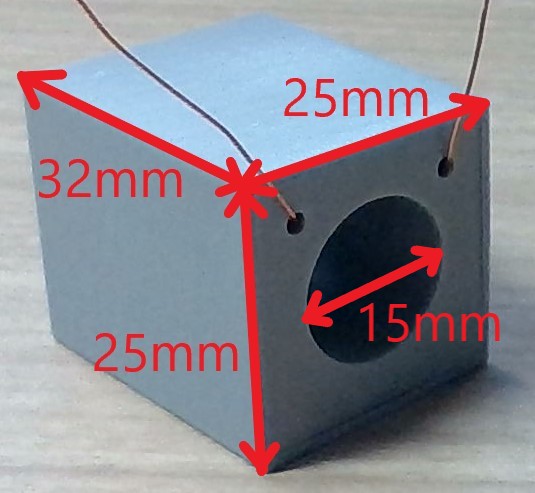}
        \subcaption{Size of coil part.}
        \label{fig2.1}
      \end{minipage} &
      \begin{minipage}[t]{0.45\hsize}
        \centering
        \includegraphics[height=30mm,width=40mm]{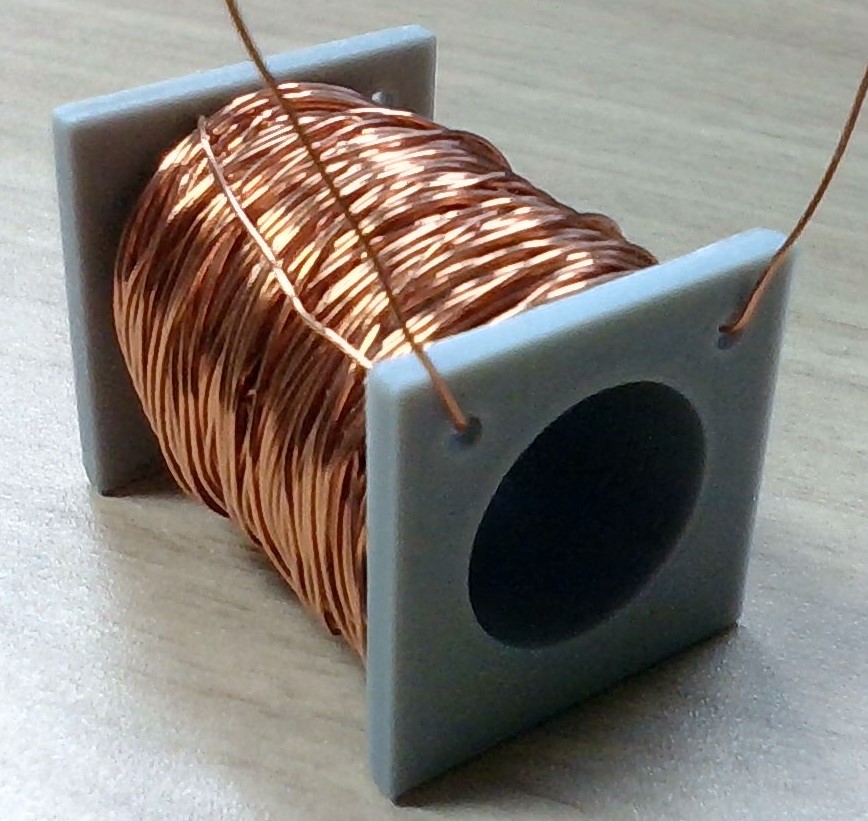}
        \subcaption{Inside of coil part.}
        \label{fig2.2}
      \end{minipage} \\
   
      \begin{minipage}[t]{0.45\hsize}
        \centering
        \includegraphics[height=30mm,width=40mm]{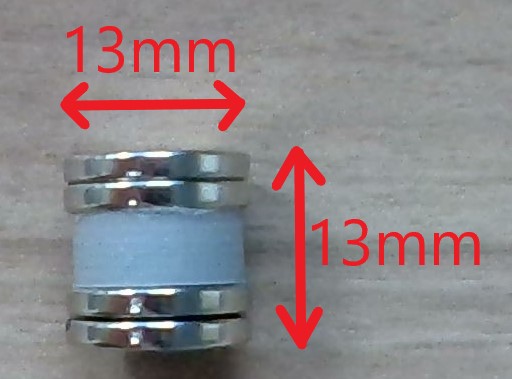}
        \subcaption{Size of magnet part.}
        \label{fig2.3}
      \end{minipage} &
      \begin{minipage}[t]{0.45\hsize}
        \centering
        \includegraphics[height=30mm,width=40mm]{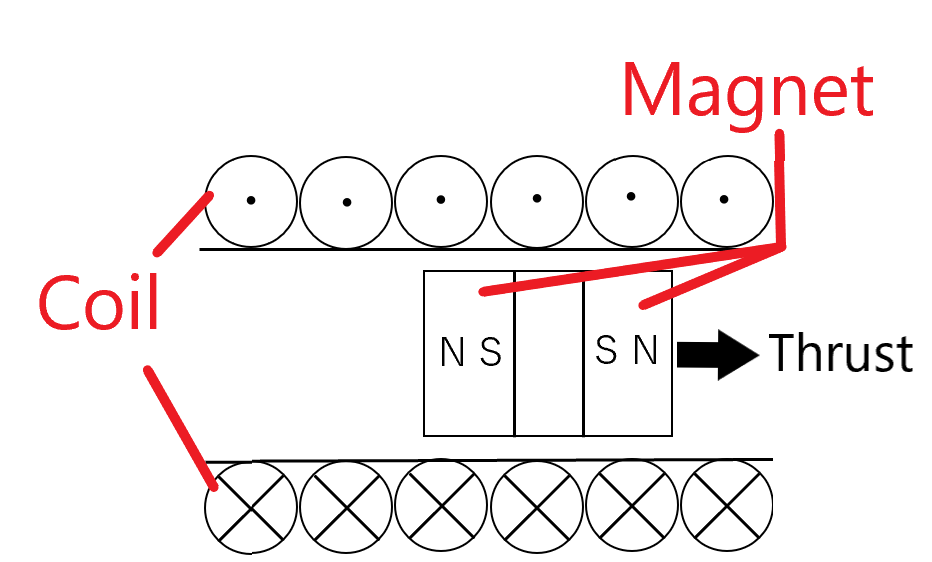}
        \subcaption{Driving principle.}
        \label{fig2.4}
      \end{minipage} 
    \end{tabular}
    \caption{The proposed actuator.}
  \end{figure}

The driving principle of the actuator used in the proposed device is that a magnet is placed inside a coil and an electric current is applied to the coil to generate thrust due to Lorentz force, which causes the magnet to move in a straight line, as shown in Figure 3(d).
The actuator is attached to the back of the hand, and a thread is attached to the magnet. Velcro tape is attached to the end of the string and secured to the finger. When the actuator begins a linear motion, the finger is pulled (Figure 4).
A 2A current is applied to each coil to activate the actuator. The actuators are electrically controlled by a microcontroller (Arduino Nano Every), a transistor (2SC3422), a 300 resistor, and a general-purpose rectifier diode (1N4007G).

\begin{figure}[htbp]
    \begin{tabular}{cc}
      \begin{minipage}[t]{0.45\hsize}
        \centering
        \includegraphics[height=30mm,width=40mm]{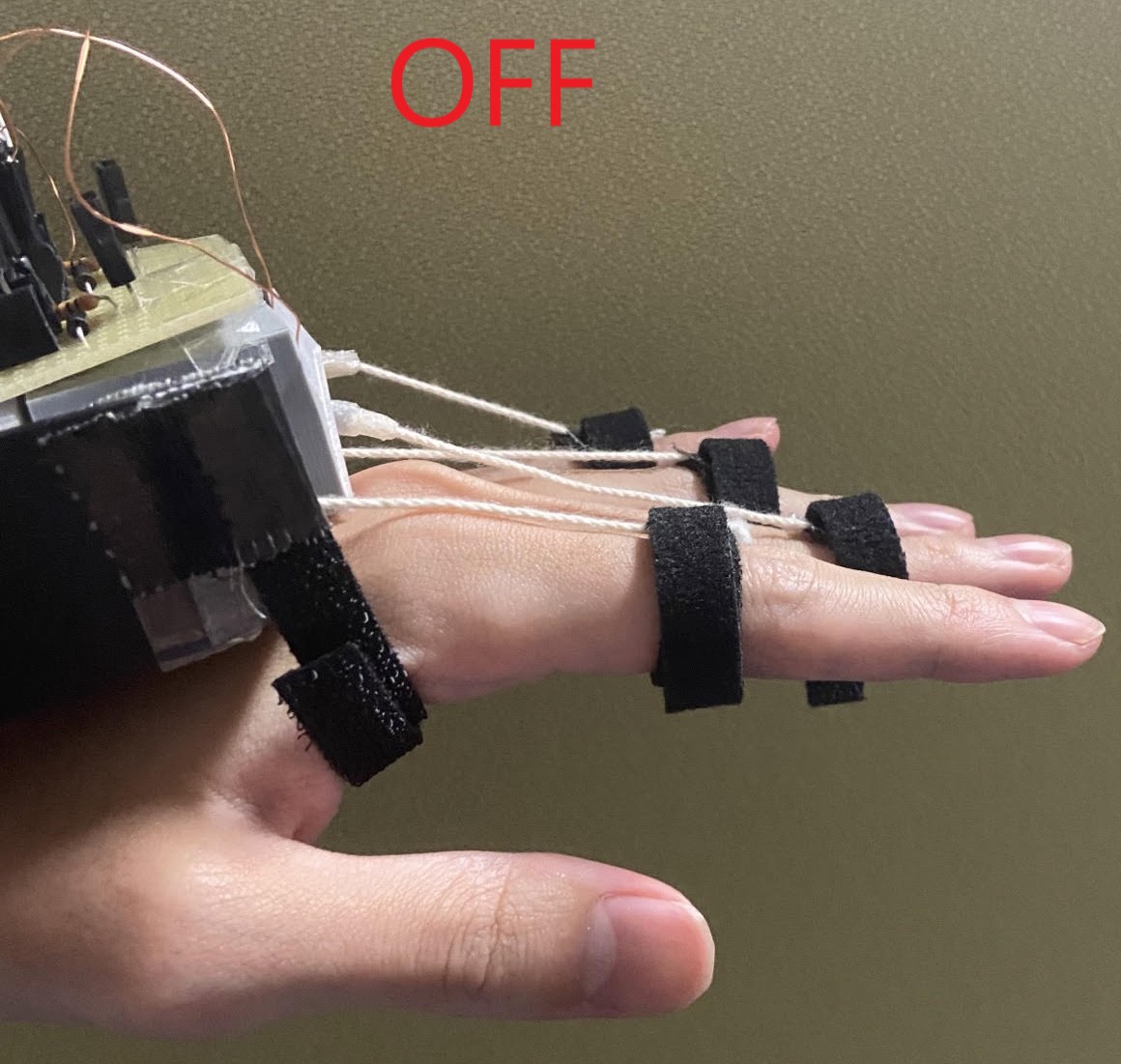}
        \label{3.1}
      \end{minipage} &
      \begin{minipage}[t]{0.45\hsize}
        \centering
        \includegraphics[height=30mm,width=40mm]{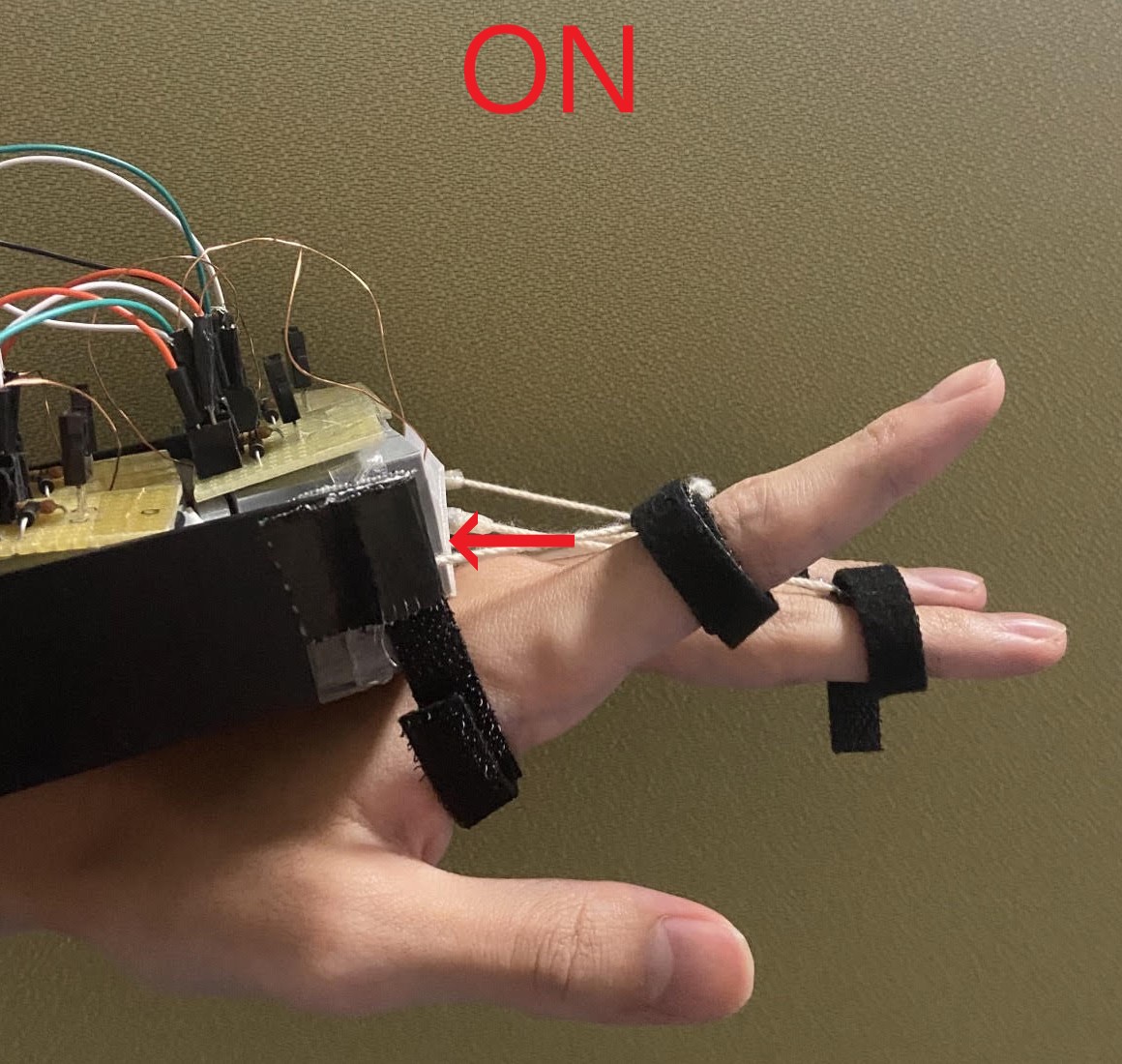}
        \label{3.2}
      \end{minipage}
    \end{tabular}
    \caption{The actuator under ON and OFF states.}
  \end{figure}

\section{Evaluation experiment}

To confirm the usefulness of the proposed device, we conducted an evaluation experiment of the proposed system. Five graduate students (aged 22 to 26 years old, males) were asked to participate in the experiment. For the questionnaire, a five-point Likert scale (5 denotes “strongly agree" and 1 for “strongly disagree") was used to determine the subjects' opinions about the questions in detail. 
The questions included are as follows:
\begin{itemize}
    \item a) Was it easy to move your fingers?
    \item b) Was it easy to understand the system in a short time?
    \item c) Was it easy to put the device on?
    \item d) Was it easy to feel the force exerted on the fingers?
    \item e) Is it useful for learning movements?
    \item f) Are you satisfied with the finger guidance?
\end{itemize}

\subsection{Finger Force-Sensing Task}

In the experiment E1 for finger force-sensing tasks, we let the subjects determine which actuator works without telling them which actuator really worked, and let the subjects determine which actuator works. The proposed device was attached to the subjects' left hands, and the subjects noted which finger applied the force. We determined the specific finger movements in advance, as shown in Figure 5(a). When the system was activated, the actuator automatically operated, and the force was applied to the fingers. We considered the movements of four fingers: index, middle, ring, and little finger. We set the force to not only one finger but up to three fingers at the same time. There was an interval of 5 seconds between the first task of force to a finger and the second task, and the subjects were asked to distinguish which finger was guided during the task. This procedure was repeated five times. The subjects were asked to not look at the device on their left hand during the experiment. Figure 5(b) shows a photo of a participant in our experiment.

\begin{figure}[htbp]
 \centering
  \begin{minipage}[b]{0.65\linewidth}
    \includegraphics[width=60mm]{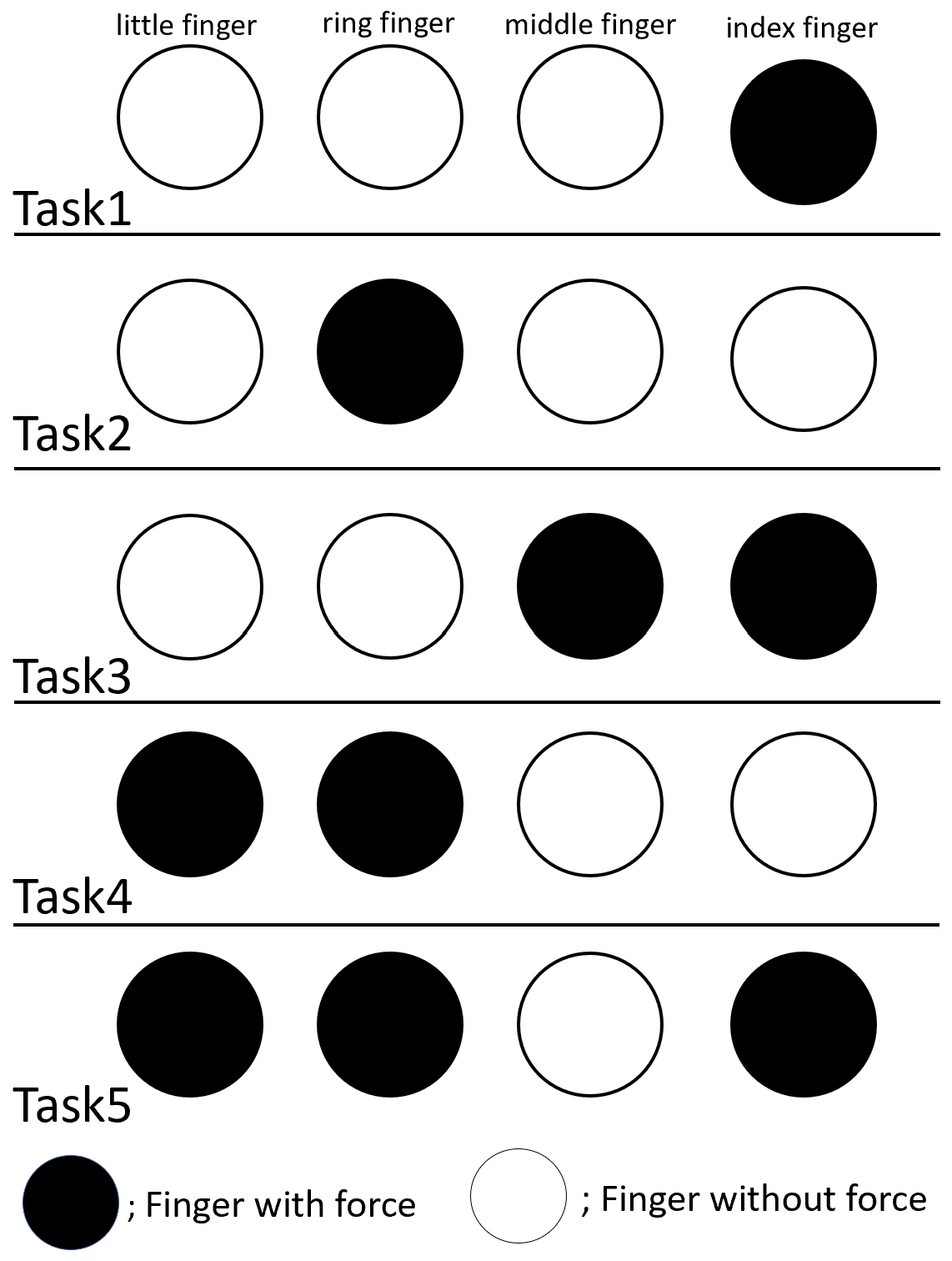}
    \subcaption{Tasks for finger force-sensing.}
  \end{minipage}
  \begin{minipage}[b]{0.65\linewidth}
    \includegraphics[width=60mm]{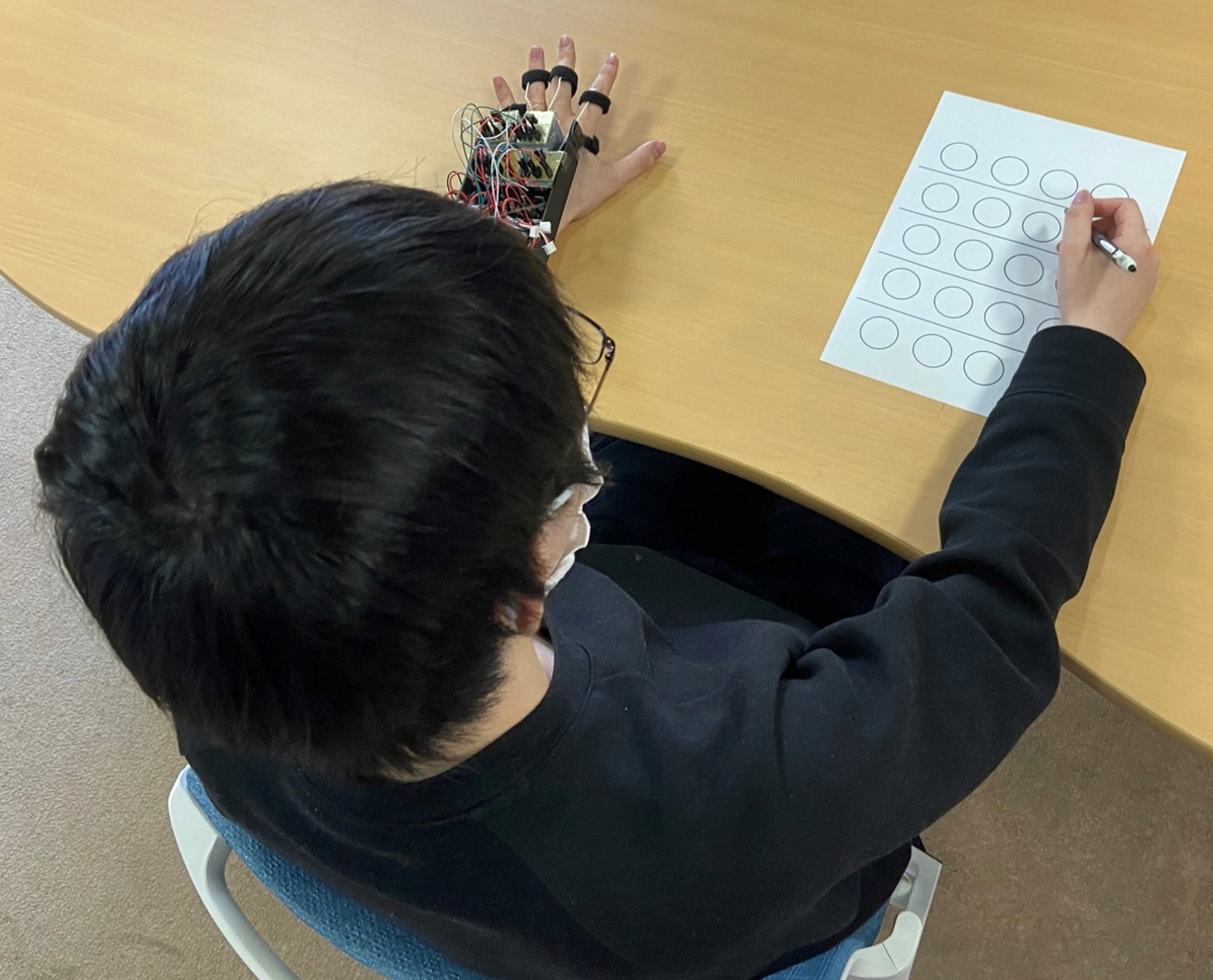}
    \subcaption{Photo of a participant in our experiment.}
  \end{minipage}
  \caption{The evaluation experiment of MagGlove.}
\end{figure}

\subsection{Force-Sensing with Different Time Intervals}

In the experiment E2 for force-sensing with different time intervals, the subjects were notified that a force would be applied for 0.3 or 3.0 seconds in advance. The subjects were asked to flex the finger if they felt the force was applied for 0.3 seconds and to extend the finger if they felt the force was applied for 3.0 seconds. We applied a force to the index, middle, ring, and little fingers for either 0.3 or 3.0 seconds. The order in which the forces were applied to the fingers was not told to the subjects. There was an interval of 3.0 seconds between the first application of force to a finger and the second application of force to a finger.  
In experiment E1, we set the time interval between re-stimulation to 5 seconds to have the participants note which finger the force was applied to, but in experiment E2, we decided that the interval should be shorter than in the first experiment because the participants only had to extend or flex the finger to which the force was applied. 
In this experiment, the program determined which finger would be subjected to the force and the time in advance. In the actual experiment, the index finger had applied force for 3.0 seconds, the middle finger for 0.3 seconds, the ring finger for 3.0 seconds, and the little finger for 0.3 seconds in this order.

\section{Results and Discussion}

\subsection{Finger Force Sensing}

Figure 6 shows the results of experiment E1. All five subjects were able to discriminate between Task 1 and Task 2, as shown in Figure 5(a), in which only one finger was used to apply force; however, in Tasks 3 and 4, in which two fingers were used, only three and four of five subjects could discriminate between the two tasks, respectively. In Task 5 in which three fingers were used to apply force, only one of five subjects were able to discriminate between them. We found that as the number of fingers to which force was applied increased, the tendency was for the subjects to become less able to recognize to which finger force was applied.

\begin{figure}[htbp]
\begin{center}
\centerline{\includegraphics[height=40mm,width=80mm]{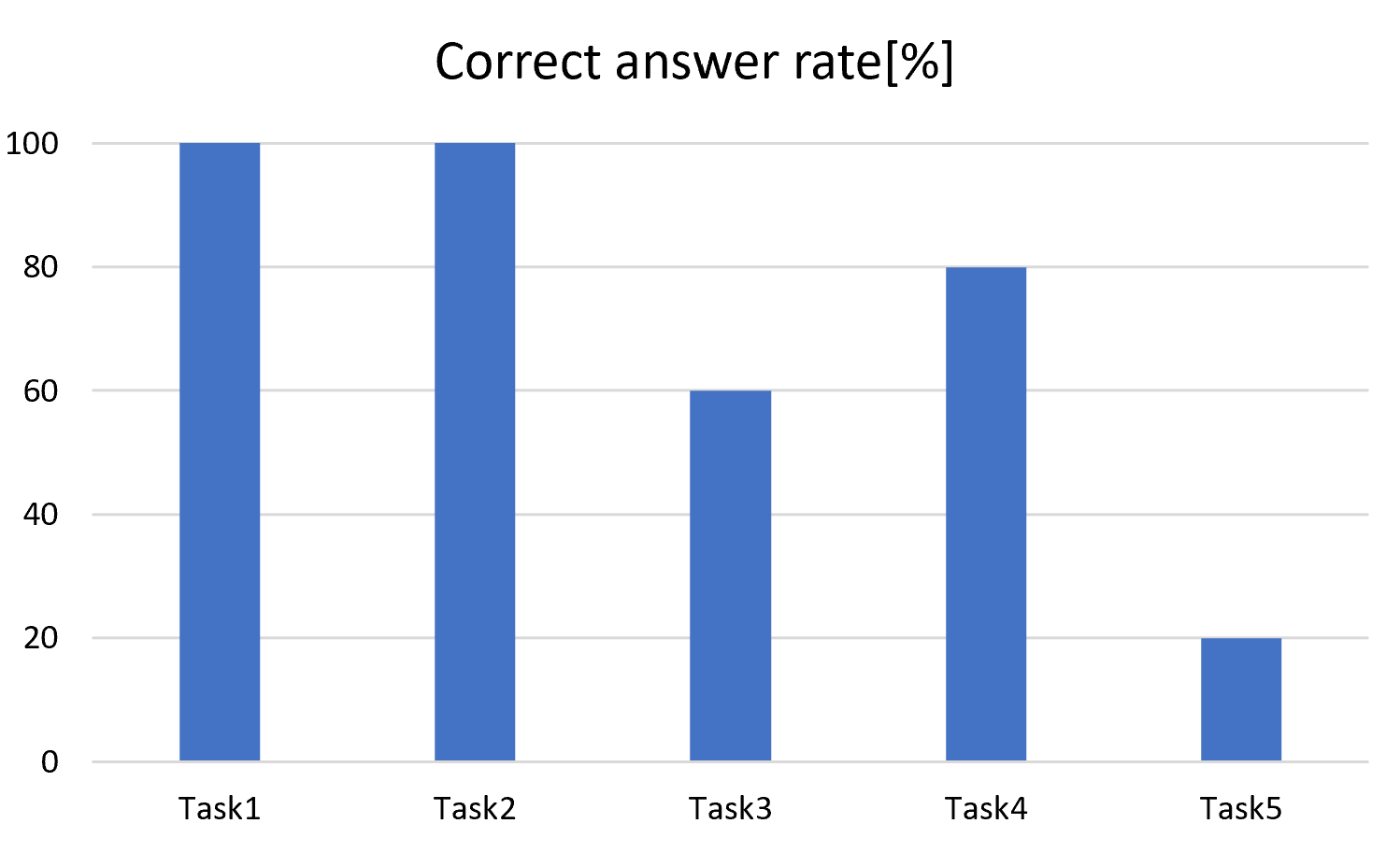}}
\caption{Results of finger discrimination experiment.}
\label{fig5}
\end{center}
\end{figure}

\subsection{Time Intervals}

As a result of experiment E2, four of five subjects were able to perform the finger extension/flexion correctly. On the other hand, one of the subjects made a mistake in finger extension/flexion. This is not because the subject could not distinguish the time when the force was applied but because the force applied to the finger was too weak to be recognized. For the proposed MagGlove device, the same actuator is applied to all fingers, and the magnitude of applied force is almost constant. Therefore, the reason that only the index finger could not be recognized may be related to the way the device is attached. It is difficult to recognize the force unless the device is attached when the thread is fully stretched to some extent. Although the finger length differs from person to person, the length of the thread is fixed. Therefore, the device may not be effective depending on the length of the fingers and the place where the thread is fixed.

\begin{figure}[htbp]
\begin{center}
\centerline{\includegraphics[height=50mm,width=90mm]{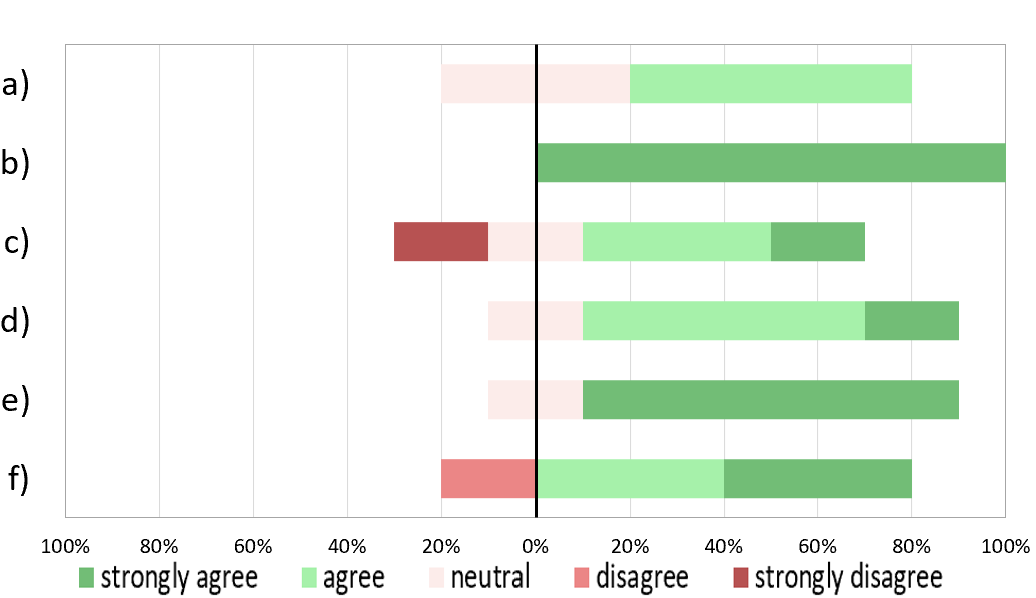}}
\caption{Evaluation result of the questionnaire. }
\label{fig6}
\end{center}
\end{figure}

The results of the questionnaire survey are shown in Figure 7. For question a), some participants reported that it was easy to move the fingers, while others were undecided. This is because the thread may get caught when flexing the fingers. The length of thread is the same, so the ease of finger movement is considered to depend on the length of the subject's fingers and the way the thread is attached. For question b), all five subjects gave a rating of 5. Therefore, there is no problem with the proposed device system in relation to question b) . On the other hand, for question c), the ease of wearing the device was rated differently by the subjects. One reason is that the device is too large, and it takes time to put it on when the subject cannot use one hand. Another reason is that the Velcro tape used to secure the thread to the finger sticks to the Velcro tape on the other fingers, making it difficult to apply the device. 
 
 \subsection{Expected Applications}
 
 The proposed device has potential applications in various activities, such as instrumental practice (Figure 8(a)). Another promising application is haptic sensing in VR and AR environments, as shown in Figure 8(b). When a user tries to touch virtual objects, an actuator is activated to generate a drag force to present haptics to the user. In this work, the purpose of the proposed device is for human augmentation to support the learning of finger motion by extending and bending the fingers, but it could also be used for guidance, such as navigation by changing the fingers to which the force is applied and the time the force is applied to those fingers.
 
\begin{figure}[htbp]
  \begin{minipage}[b]{0.45\linewidth}
    \centering
    \includegraphics[height=30mm,width=40mm]{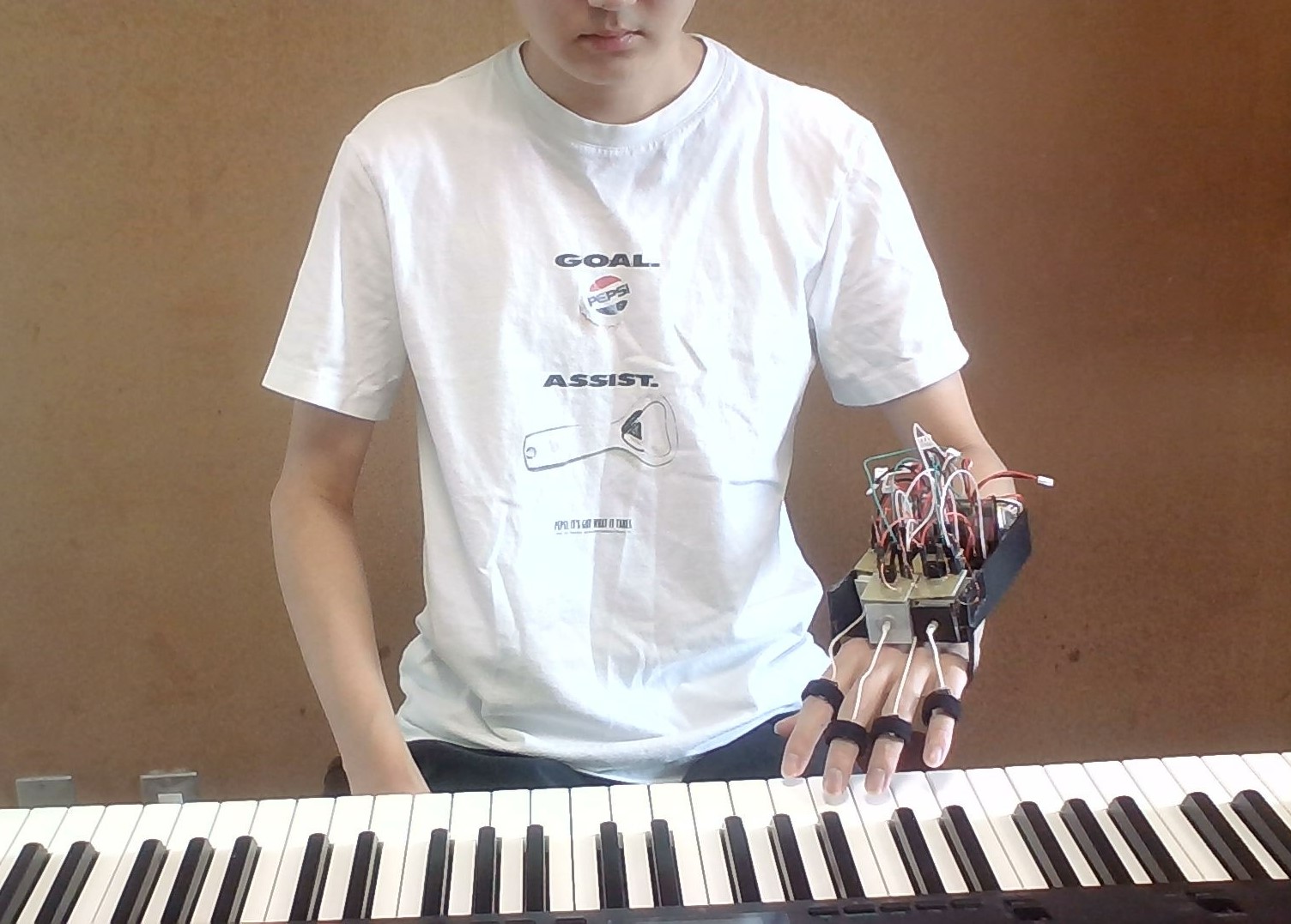}
    \subcaption{Piano practice.}
  \end{minipage}
  \begin{minipage}[b]{0.45\linewidth}
    \centering
    \includegraphics[height=30mm,width=40mm]{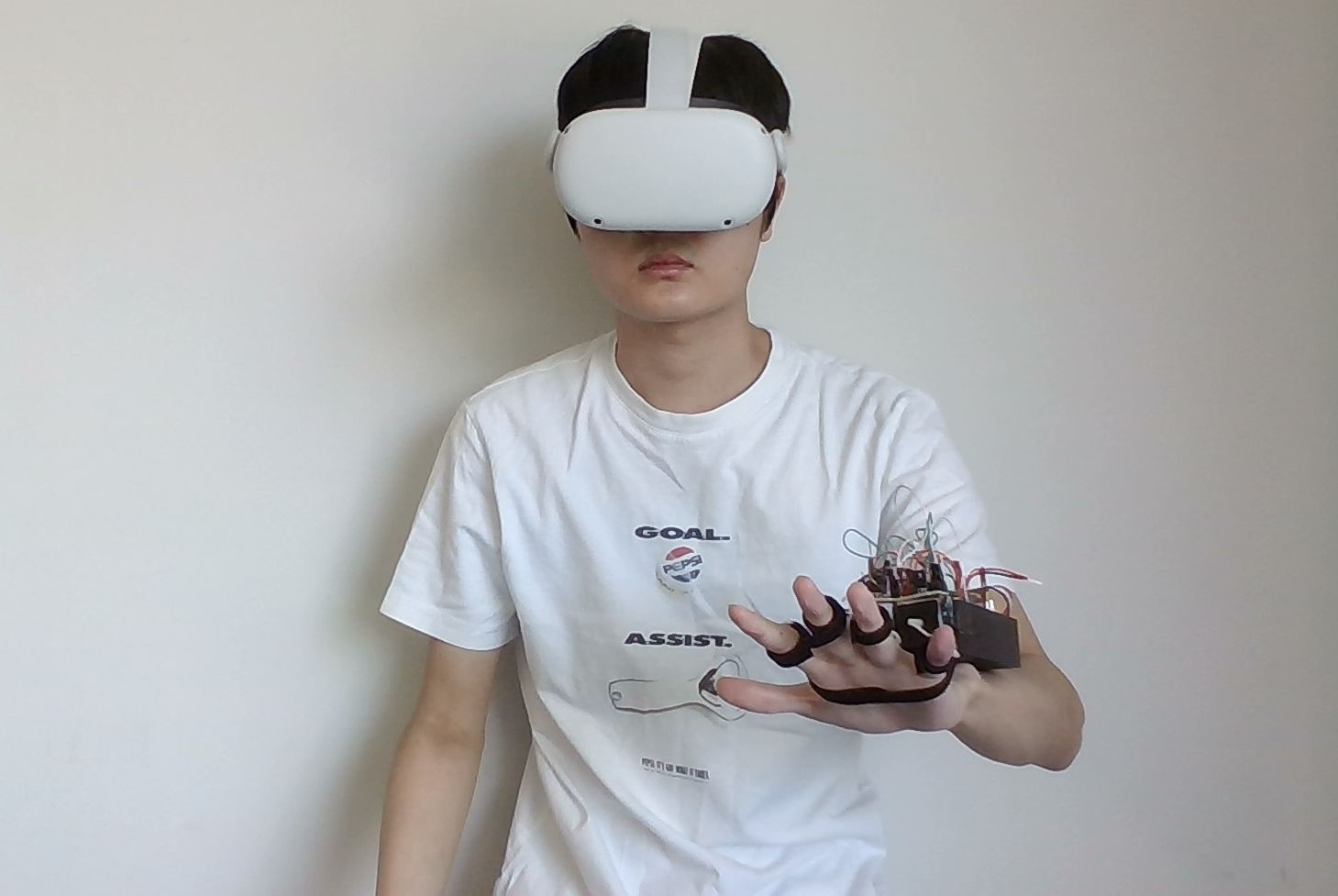}
    \subcaption{VR.}
  \end{minipage}
  \caption{Example of expected applications.}
\end{figure}

\section{Conclusion}

In this work, we proposed a haptic glove, MagGlove, with a movable magnet for manipulation learning to augment users' capability in finger movements. We conducted two experiments to evaluate the usefulness of the proposed device. From the experiment results, we confirmed the usefulness and shortcomings of the proposed method. We found that the subjects were able to perceive the force presentation of the device in part, though not in some cases when the conditions were complex. Therefore, it is verified that the proposed MagGlove device is useful for simple manipulation learning tasks, but it could be improved for  complex tasks, such as rapid movements of multiple fingers.

We plan to increase the number of participant in the experiment in the future. During the experiments, some subjects could perceive the same forces, while others were not able to perceive them. Therefore, it is necessary to calibrate the device to ask the users to complete several tasks to determine the perception levels, and then to set the appropriate force for each user accordingly. The other possible solution is to adjust the amount of force applied to the fingers by changing the value of the electric current applied to the actuator depending on the stage of manipulation learning. For example, when learning to play a musical instrument, the force applied to the fingers could be increased for beginners to make it easier to perceive. For advanced players, the device can decrease the force level to make it harder to perceive. We are also considering improving the actuator by using a neodymium magnet with a higher magnetic flux density and a smaller size than the neodymium magnet used in the current prototype. This improvement may reduce the burden on the user so that the user will not feel fatigued even after a long time wearing the device.

\section*{Acknowledgements}
We thank all participants in our experiments and anonymous reviewers. This work was partially supported by JAIST Research Grant, and JSPS KAKENHI grant JP20K19845, Japan. 

\bibliographystyle{unsrt}
\bibliography{bibliography.bib}

\end{document}